\documentclass{iau}
\usepackage{graphicx} 

\title[IAUS291.~~Compactness of isolated neutron stars] %% short title %%
{Constraints of the compactness of the isolated neutron stars via X-ray phase-resolved spectroscopy} %% full title %%

\author[V. Hambaryan et al.]  %% short author list %%
{V.\,Hambaryan$^1$,
 V.\,Suleimanov$^2$,
 R.\,Neuh\"auser$^1$
\and K.\,Werner$^2$
}

\affiliation{$^1$Friedrich-Schiller-Universit\"at, Astrophysikalisches Institut und Universit\"ats-Sternwarte, \\07745 Jena, Germany, \\  email: {\tt valeri.Hambaryan@uni-jen.de} \\[\affilskip]
$^2$Eberhard-Karls-Universit\"at, Institut f\"ur Astronomie und Astrophysik, \\72076 T\"ubingen, Germany}

\pubyear{2012}
\volume{291}  %% insert here IAU Symposium No.
\jname{\mbox{Neutron Stars and Pulsars: Challenges and Opportunities after 80 years}}
\editors{J. van Leeuwen, ed.} 
\begin{document}

\maketitle

%% -- Abstract ----------------------------------
\begin{abstract}
 A model with a condensed iron surface and partially ionized hydrogen-thin atmosphere allows us to 
fit simultaneously the observed general spectral shape and the broad absorption feature (observed at 0.3 keV) 
in different spin phases of the isolated neutron star RBS 1223. We constrain some physical properties of the 
X-ray emitting areas, i.e. the temperatures ($\mathrm{T}_{pole1} \sim \mathrm{105\,eV, T}_{pole2} \sim \mathrm{99\,eV}$), 
magnetic field strengths 
($\mathrm{B}_{pole1} \approx \mathrm{B}_{pole2} \sim \mathrm{8.6 \times 10^{13} \,G}$) at the poles, and their distribution parameters 
(a1 $\sim$ 0.61, a2 $\sim$ 0.29, indicating an absence of strong toroidal magnetic field component). 
In addition, we are able to place some constraints on the geometry of the emerging X-ray emission and 
the gravitational redshift ($\mathrm{z}\sim\mathrm{0.16^{0.03}_{-0.01}}$) of the isolated neutron star RBS 1223.
\end{abstract}

% add below any authors, subjects and objects for indexing 
%   add more lines if necessary
%   but leave all lines commented out
%\index[author]{LastName1, Initials|textbf}
%\index[author]{LastName2, Initials|textbf}
%\index[subject]{Keyword1}
%\index[subject]{Keyword2}
%\index[object]{Object1}
%\index[object]{Object2}

\firstsection % if your document starts with a section,
\section{Introduction}

Observations and modeling of thermal emission from isolated neutron stars 
can provide not only information on the physical properties such as the magnetic field, temperature, 
and chemical composition of the regions where this radiation is produced, but also 
we may infer on the properties of matter at higher densities deeper inside the star. 

In particular, the study of thermal emission from isolated neutron stars may allow one to infer 
the surface temperature and total flux measured by a distant observer and to estimate the real 
parameters. With the known distance  and the redshifted radius of the neutron star 
the actual radius and mass of a neutron star are:
\centerline{$R = R^\infty [1-2 G M/R c^2]^{1/2},\,\,\,
M = \frac{c^2 R}{2 G} \left[ 1 - \left(\frac{R}{R^\infty}\right)^2 
\right]\,\,\,$ (see, e.g. \cite[Zavlin (2009)]{zv1})}.

The detection and identification of any absorption/emission
feature in the spectrum or performing rotational phase-resolved spectroscopy of
isolated neutron stars will allow us to determine gravitational redshift and directly
estimate the mass-to-radius ratio, $M/R$. Together they yield a unique solution for
$M$ and $R$. These spectral features may allow to measure
the neutron star magnetic field and provide an important input for modeling of magnetized
atmospheres. 

\begin{figure}[!h]
 \begin{center}
      \includegraphics[width=13.5cm]{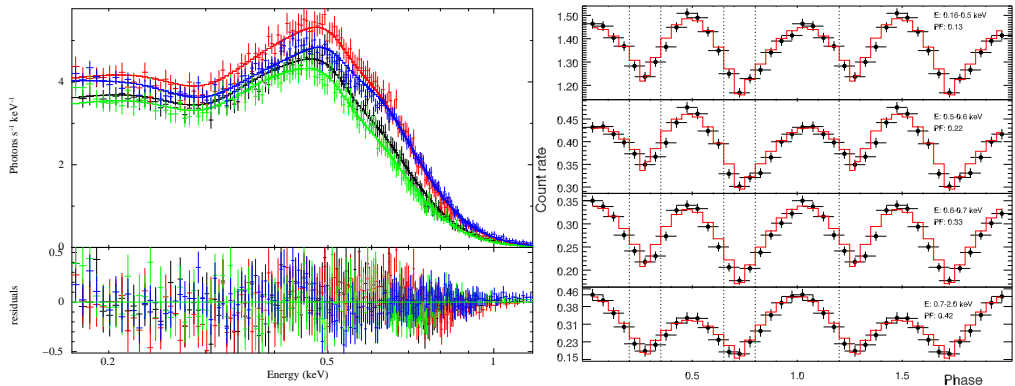}
  \caption{\emph{XMM-Newton EPIC} pn co-added phase-averaged X-ray spectra (left panel) 
 including primary and secondary peaks, first and second minima, and phase-folded light 
curves (right panel) in different energy bands of RBS 1223 combined from  12 pointed observations.}
 \label{fig1}
 \end{center}
\end{figure}

RBS 1223 shows the highest pulsed fraction (13-42\%, depending on
energy band, see Fig.~\ref{fig1}) and strongest broad absorption
feature (\cite[Schwope 
et al. 2007]{2007Ap&SS.308..619S}) of all isolated neutron stars.

High quality rotation phase resolved spectroscopy is needed in order
to fit with neutron star magnetized atmosphere models and to constrain
the gravitational redshift of the neutron star (\cite[Suleimanov et al. 2010; Hambaryan et al. 2011]{vsvh,vhvs}).

\section{Data analysis and results}

Using the data collected with \emph{XMM-Newton} EPIC pn  
from the 12 publicly (similar instrumental setup, i.e. Full Frame, Thin1 Filter) available observations, 
in total presenting about 175~ks of effective exposure time, 
we extracted spin-phase resolved spectra with high S/N ratio (Fig.~\ref{fig1}) and fitted
simultaneously with highly magnetized isolated neutron star surface/atmosphere models (\cite[Suleimanov et al. 2010]{vsvh}).
These models are based on various local models and compute rotational phase 
dependent integral emergent spectra of isolated neutron star, 
using analytical approximations.

 The basic model includes temperature/magnetic field distributions over isolated 
neutron star surface\footnote{$T^4 = T_\mathrm{ p1,2}^4 \frac{\cos^2\theta}{\cos^2\theta + a_\mathrm{ 1,2}\sin^2\theta } + T_\mathrm{ min}^4, B = B_\mathrm{ p1,2} \sqrt{\cos^2\theta + a_{1,2}\sin^2\theta}$, where the parameters $a_\mathrm{ 1,2}$ are approximately equal to the squared ratio of the magnetic
field strength at the equator to the field strength at the pole,
$a_\mathrm{ 1,2} \approx (B_\mathrm{ eq}/B_\mathrm{ p1,2})^2 $.
Using these parameters we can describe various temperature distributions, from strongly 
peaked ($a \gg 1$) to the classical dipolar ($a = 1/4$)  and homogeneous ($a = 0$) ones.},
viewing geometry and gravitational redshift. Three local
radiating surface models are also considered, namely, a naked condensed iron surface (\cite[van Adelsberg \& Lai 2006]{2006MNRAS.373.1495V}) and 
partially ionized hydrogen model atmospheres,  
semi-infinite or finite atop of the iron condensed surface. 

\begin{figure}[!h]
  \begin{minipage}[b]{0.65\textwidth}
    \includegraphics[width=\textwidth]{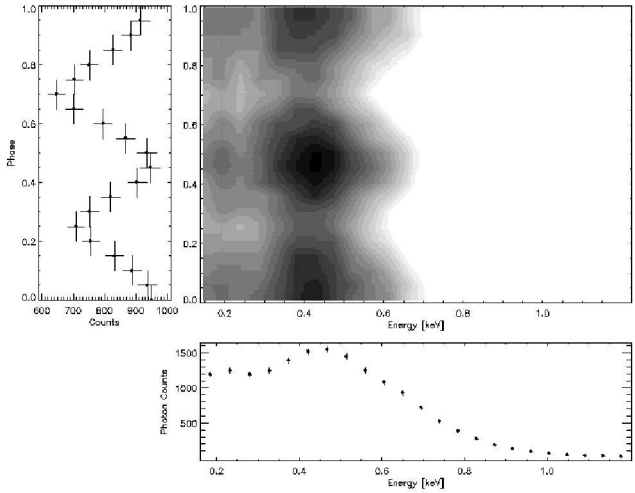}\vspace{-5mm}%
  \end{minipage}\hfill
  \begin{minipage}[b]{0.33\textwidth}
    \caption{Energy-Phase image of RBS1223 combined from different observations. Rotational phase-folded  
light curve in the broad energy, 0.2-2.0~keV, band (left panel) and the phase averaged spectrum (bottom panel) 
are shown.} \label{fig2}
  \end{minipage}
\end{figure}
\begin{figure}[!h]
%\vspace*{-5.0cm}
 \begin{center}
  \includegraphics[width=13.5cm]{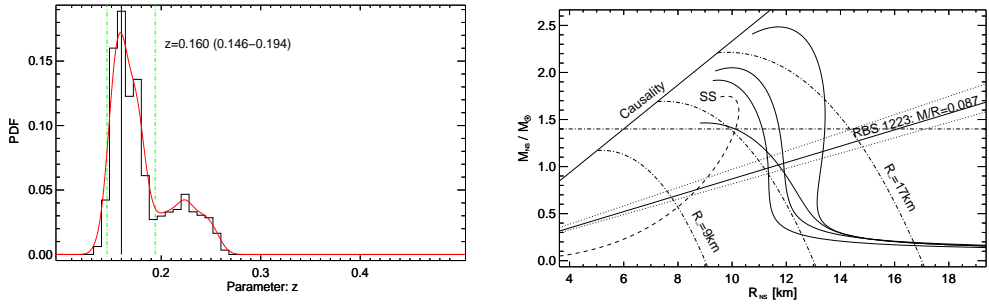}
  \caption{Probability density distribution  of gravitational redshift by 
Markov Chain Monte Carlo (MCMC) fitting with the model of 
strongly magnetized neutron star condensed Iron surface and partially ionized Hydrogen thin atmosphere atop it (left panel).
Dashed vertical lines indicate the highest probability interval ($68\%$).
 Mass-radius relations (right panel) for several EOS (\cite{2007ASSL..326.....H}, thin solid
   curves), and a strange star (thin dash-dotted).  Thick dashed: curve of
   constant $R^\infty=R/\sqrt{1-2GM/Rc^2}=17$ km (\cite{2005esns.conf..117T}. }
  \label{fig3}
 \end{center}
\end{figure}

The observed phase resolved spectra (i.e. energy-spin phase image, Fig.~\ref{fig2}) of the isolated neutron star RBS 1223   
are satisfactorily fitted (verified also via MCMC, Fig.~\ref{fig3}) with two slightly different physical and geometrical  
characteristics of emitting areas, a model parameterized with a Gaussian 
absorption line superimposed on a  blackbody spectrum and by 
the model of a condensed iron surface, with partially ionized,  
optically thin hydrogen atmosphere above it, including vacuum polarization effects,  
as orthogonal rotator. Note, the latter one is more physically motivated.
We have additionally performed Markov Chain Monte Carlo (MCMC) fitting 
as implemented in {\it XSPEC}. 

The fit also suggests the absence of a strong toroidal  
magnetic field component. Moreover, the determined mass-radius ratio,
$(M/M_{Sun})/(R/\mathrm{km})=0.087 \pm 0.004 $, 
suggests a very stiff equation of state of RBS 1223. 

\begin{acknowledgment}
 We acknowledge support by the German
      \emph{Deut\-sche For\-schungs\-ge\-mein\-schaft (DFG)\/} through project
      C7 of SFB/TR~7 ``Gravitationswellenastronomie'' 
\end{acknowledgment}

\end{document}